# Strategy of a separation technique for different particles with the same size and zeta potential: Application of non-additive Asakura-Oosawa theory

Ikuma Ogasawara[a] and Ken-ichi Amano[a*]

[a] *Faculty of Agriculture, Meijo University, Nagoya, Aichi 468-8502 , Japan .*

[*] Correspondence author: K. Amano (amanok@meijo-u.ac.jp)

## ABSTRACT

In this letter, we use knowledge gained from our recent study to present a technique for separation of nanoparticles such as exosomes, anticancer drugs, and vaccines. The technique involves adding non-adsorptive polymers to a system in which two types of nanoparticles with the same size and zeta potential are dispersed. The different types of nanoparticles can be separated based on differences in their hydrophobicities and softness of the polymer. Using the non-additive Asakura-Oosawa theory and assuming a realistic model system, we were able to separate the two types of the particles with the same size and zeta potential in the model system.

## MAIN TEXT

Separation techniques such as ultracentrifugation by T. Svedberg [1], electrophoresis by A. Tiselius [2], and partition chromatography by A. J. P. Martin and R. L. M. Synge [3] have been developed. All of these separation techniques have been awarded the Nobel Prize in Chemistry. These separation techniques have been used in a wide range of fields, and further developments of them are very important.

Cells secrete exosomes, vesicles that deliver chemical information between cells [4], and are attracting attention as new biomarkers [5], regenerative medicine [6], and drug delivery system (DDS) capsules [7]. However, since the exosomes are secreted from various cells and have different sizes and constituent components such as membrane proteins and polysaccharides, it is necessary to separate and purify the exosomes before they are used practically. Ultracentrifugation and polymer-based precipitation methods [8] are mainly used to isolate the exosomes. In the study by Bibette [9], he added micelles instead of the non-adsorbent polymers in the polymer-based precipitation method into the O/W emulsion, which caused flocculation of larger micro oil droplets. Then, creaming of the flocs occurred and the other droplets were remained in the bulk emulsion, leading to size separation of the micro oil droplets. In the micelle-based separation, Asakura-Oosawa (AO) theory [10] was applied. The AO theory explains that the larger the size of particles are, the larger overlap volumes of the excluded volumes of the particles become, which causes aggregation of the larger particles.

These separation methods above utilize differences in particle size and/or zeta potential, which cannot separate exosomes of the same size and zeta potential. In nature, there are various types of exosomes but they have similar properties. Hence, separation of the similar exosomes is one of the important challenges. In the present letter, we propose a strategy for separating different particles with the same size and zeta potential using the non-additive AO (NAO) theory [11,12].

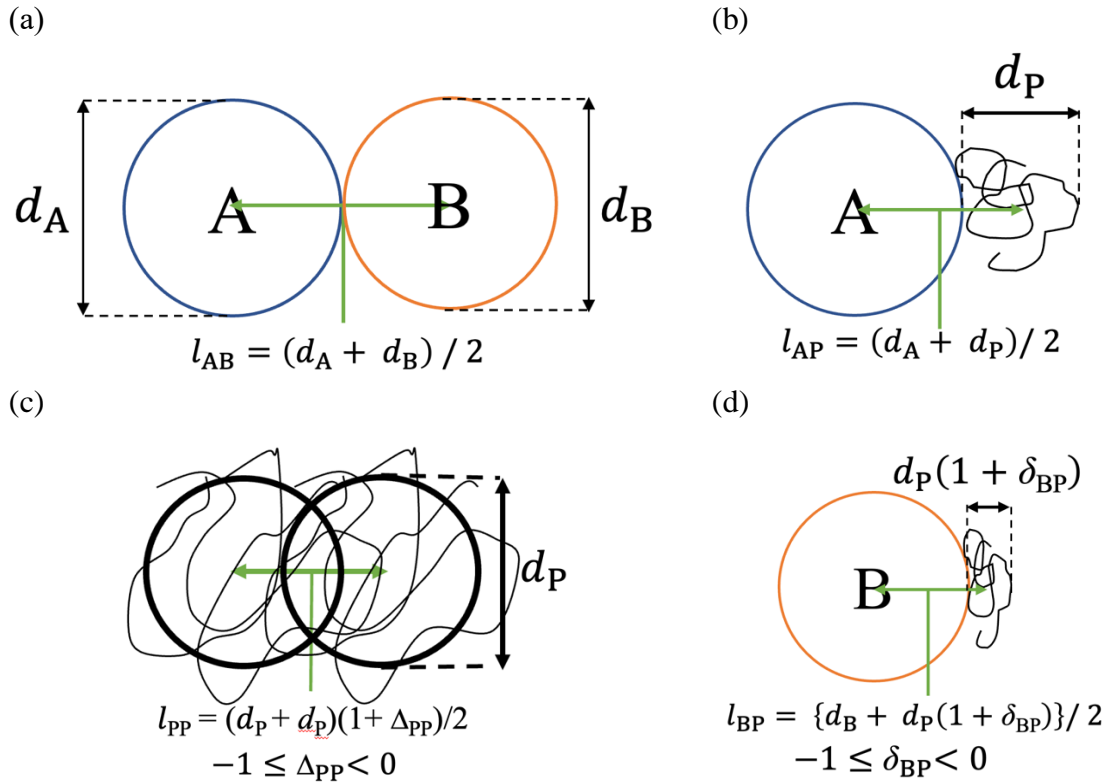

Figure 1. Schematics of contact distances $l_{AB}$ (a), $l_{PP}$ (b), $l_{AP}$ (c), and $l_{BP}$ (d).

Aim of this paragraph is explanation of the NAO theory [8]. In the general additive model (Figures 1(a) and 1(b)), the distance $l_{ij}$ between the centers of particles $i$ (= A) and particles or polymer $j$ (= B or P) is given by

$$l_{ij} = \frac{(d_i + d_j)}{2}. \tag{1}$$

Here, $d_i$ and $d_j$ are the diameters of the particles $i$ and $j$, respectively (Figure 1). Eq. (1) expresses the contact distance in the additive rule. In the NAO theory, the non-additivities in terms of the particle sizes are introduced. Since the polymers are soft, they can overlap each other (Figure 1(c)). Hence, the contact distance between the polymers can be calculated with the non-additive rule. The contact distance between the centers of polymers being $l_{PP}$ is given by using the nonadditive parameter $\Delta_{PP}$ as follows:

$$l_{PP} = d_P(1 + \Delta_{PP}) \quad (-1 \leq \Delta_{PP} < 0). \tag{2}$$

Finally, we consider a pair of the particle B and the polymer P (Figure 1(d)). We assume that only the particle B has non-additivity with the polymer P. The contact distance between them ($l_{BP}$) is expressed as

$$l_{BP} = \frac{\{d_B + d_P(1 + \delta_{BP})\}}{2} \quad (-1 \leq \delta_{BP} < 0), \tag{3}$$

Where $\delta_{BP}$ is a non-additive parameter solely designed for the pair of the particle B and the polymer P. The reason for preparing the specific parameter is as follows. In our calculation condition, the particles A and B are rigid. In addition, the particle B is more hydrophobic than the particle A, and the polymer has partly hydrophobic surfaces. In this case, the polymer tends to contact on the particle B surface with physical adsorption, leading to decrease in the contact distance between them (Figure 1(d)). Consequently, the normalized number density of the particle $i$ contacting on a rigid wall surface (contact density) in the $n$ component system ($n = 3$) can be described as follows:

$$g_{wi,cp} = exp\left[\sum_{j=1}^{n} \left\{\frac{2}{3}l_{ij}^3 - (z_{Wi} - z_{Wj})l_{ij}^2 + \frac{1}{3}(z_{Wi} - z_{Wj})3\right\}\pi\rho_j\right], \tag{4}$$

where $\pi$ and $\rho_j$ are the circular constant and the number density of the particle $j$ in the bulk, respectively. $z_{Wi}$ is the distance between the wall surface and the center of particle $i$, where the subscript W represents the wall. The subscript cp represents the contact point. Volume fraction of the particle $j$ being $\varphi_j$ can be expressed by using $\rho_j$ as follows:

$$\varphi_j = \frac{\pi}{6}\rho_j d_j^{3}. \tag{5}$$

This paragraph explains the separation technique by the NAO theory. There are two types of particles (A and B) which are models of the small extracellular vesicles or artificial nano capsules. The particles A and B have the same diameter and zeta potential. We consider two types of model systems in Figure 2. In one system, there is no water-soluble polymer. In the other system, there are the water-soluble polymers. Existences of the polymers increase crowding of the bulk and some of the particles are adsorbed in order to reduce the crowding. Furthermore, since hydrophobicity of the particle B is stronger than that of the particle A, the polymer can approach to the surface of the particle B more. This property makes the particles A preferentially coagulate and precipitate (Figure 2), which will be theoretically confirmed afterwards.

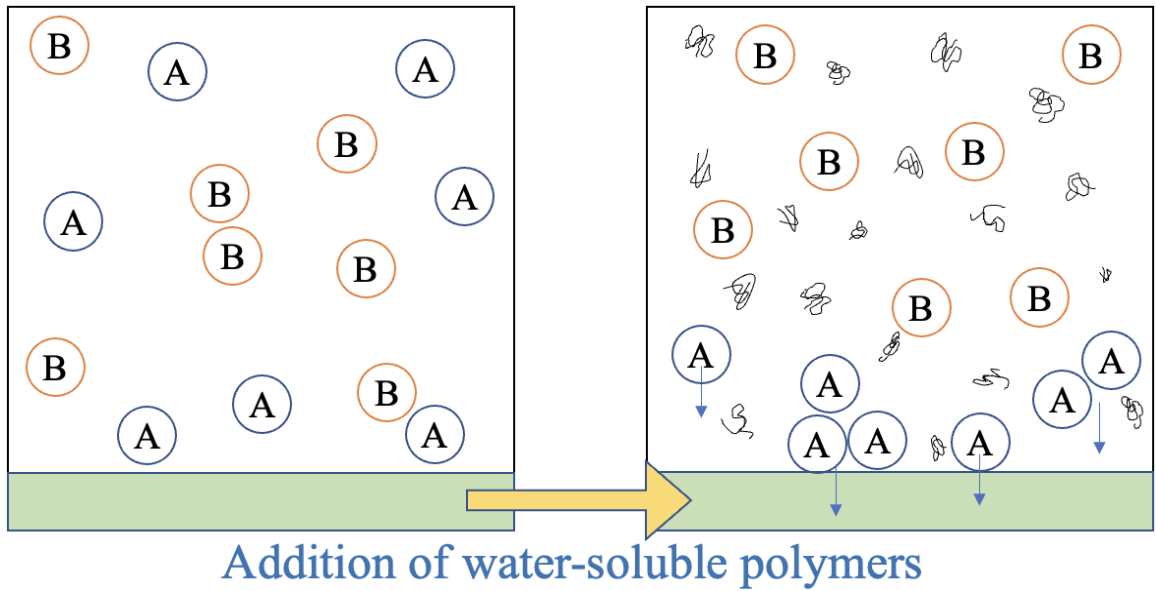

Figure 2. Illustration of the separation process of the particles A and B with the same size and zeta potential.

Applicability of the separation technique is explained by showing the calculation results of the NAO theory. The contact density of each particle vs volume fraction of the polymer is shown in Figure 3. In Figure 3, $d_A = 75$ nm, $d_B = 75$ nm, $d_P = 20$ nm, $\varphi_A =$

$1.1×10^{-5}$, $\varphi_B = 1.1×10^{-5}$, $\delta_{BP} = -0.05$, and $\Delta_{PP} = -0.15$. The higher the volume fraction of the added polymer is, the higher the contact densities of the particles A, B, and the polymer become. This phenomenon occurs, because the bulk crowding increases as amount of the added polymer increases, which facilitates them to adsorb on the wall surface, leading to decrease in the bulk crowding. Next, we compare the contact densities of the particles A and B. As shown in Figure 3, there is difference of the contact densities between them and that of the particle A is higher than that of the particle B. The difference is increased as the volume fraction of the polymer increases. This tendency is due to the difference in the contact *distances* between the particle A and the polymer and between the particle B and the polymer. This result shows that the particle A and B, which have the same size and zeta potential, can be divided depending on the affinity between the particle and the polymer.

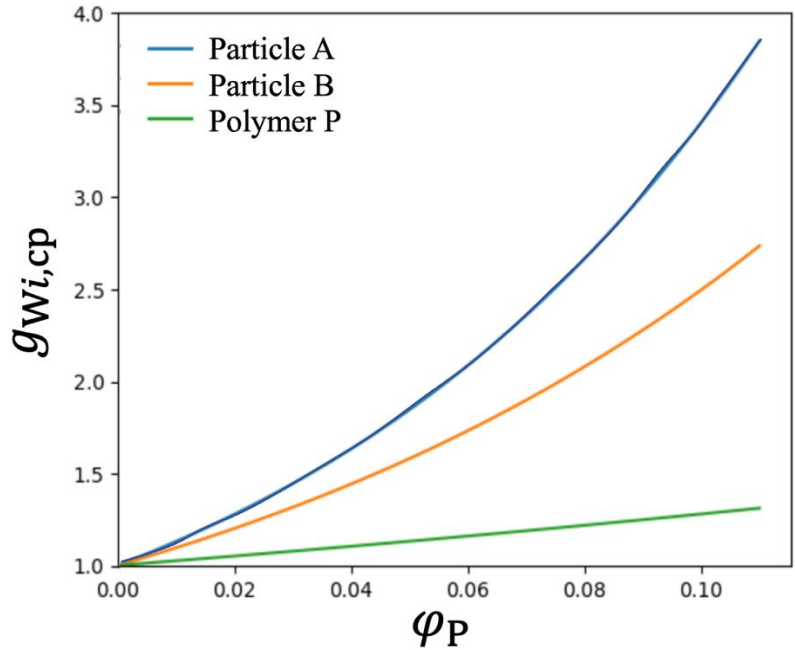

Figure 3. The contact densities of the particles A, B, and the polymer vs the volume fraction of the polymer.

The ratio of contact densities of the particle A and B ($g_{WA,cp}/g_{WB,cp}$) vs the diameter of the polymer is shown in Figure 4. In Figure 4, $d_A = 100$ nm, $d_B = 100$ nm, $\varphi_A = 1.1 × 10^{-5}$, $\varphi_B = 1.1 × 10^{-5}$, $\varphi_P = 0.1$, $\delta_{BP} = -0.05$, and $\Delta_{PP} = -0.15$. The ratio of the contact densities increases as the size of the added polymer decreases. Since the volume fraction of the added polymer is constant in our calculation, the number of polymers increases as the size becomes smaller. The increase in the number of the polymers makes the system more crowded, which is the origin of the behavior in Figure 4.

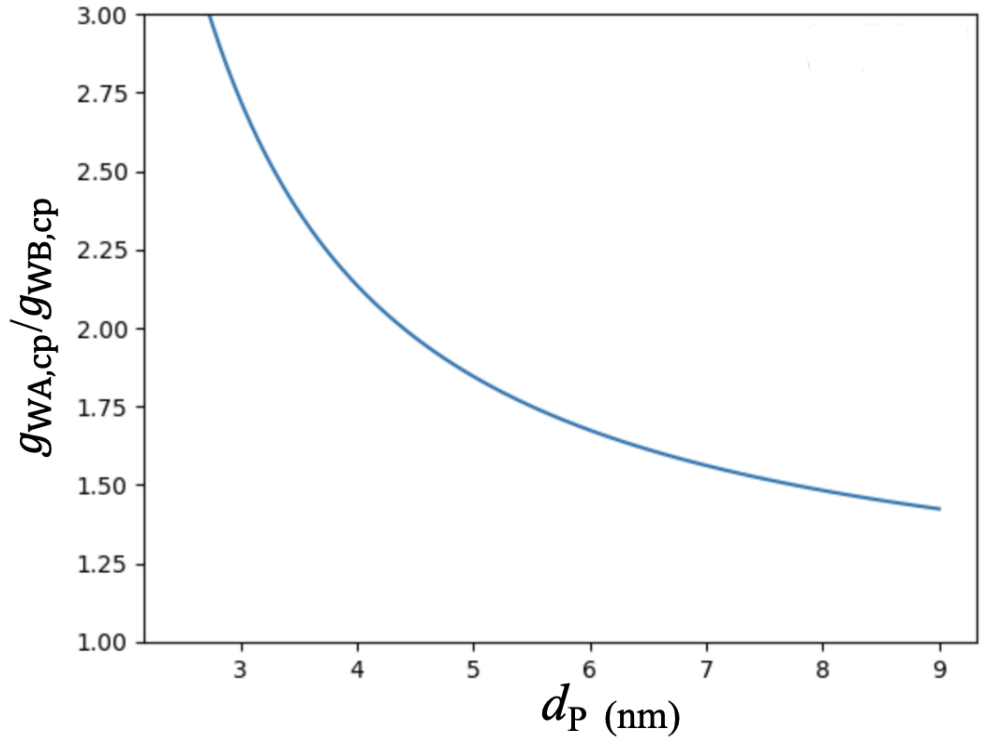

Figure 4. The ratio of the contact densities of the particles A and the particle B as a function of the diameter of the polymer. In this calculation, the volume fraction of the polymer is constant ($\varphi_P = 0.1$).

## SUMMARY

We have proposed the strategy for separating the different particles with the same size and zeta potential based on the NAO theory. Assuming the exosome mixture, we have modeled the realistic calculation condition and performed the calculation to verify the separation strategy. The calculation result has confirmed that the different particles with the same size and zeta potential can be separated. It has also found that under the condition where the volume fraction of the added polymer is constant, the smaller the size of the polymer is, the more effective separation can be achieved. This separation technique is based on the non-additivity between the particle and the polymer. The non-additivity depends on the strength of hydrophobic interaction, hydrophilic interaction, van der Waals interaction, and entropic repulsive interaction by swollen polymers between the particle and the polymer. When such an interaction force differs between the pair of the particle A and the polymer and the pair of the particle B and the polymer, the separation may be possible. Since the adsorption effect of the NAO theory is physical adsorption, the aggregates generated through the separation process can be partially dispersed later using ultrasound or a mixer. On the other hand, in the case of chemical adsorption (chemical marking), it is relatively difficult to dissociate the chemically adsorbed substance. Furthermore, the chemically marked particles tend to be easily destroyed in a process of detachment of the marked substance. For these reasons, the strategy based on the NAO theory has the potential to contribute to the separations of nano-capsules such as exosomes, vaccines, and anti-cancer drugs.

## ACKNOWLEDGEMENTS

We would like to thank Dr. R. Hanayama and Dr. T. Yoshida for useful discussion and advice.